\documentclass[fleqn,twocolumn]{article}

\newcommand{\AmS}{{\protect\the\textfont2
  A\kern-.1667em\lower.5ex\hbox{M}\kern-.125emS}}

\def\R{\rm l\!R\,}

\author
{$^1$M. V. Cougo-Pinto{\footnote{\it marcus@if.ufrj.br}}, $^1$C.
Farina{\footnote{\it farina@if.ufrj.br}} and $^{1,2}$J. F. M. Mendes{\footnote{\it jayme@if.ufrj.br}}
\\$^1${\it Instituto de F\' \i sica-UFRJ, CP 68528, Rio de Janeiro, RJ,
21.945-970}\\$^2${\it
IPD-CTEx, Av. das Am\'ericas 28.705, Rio de Janeiro, RJ, 23.020-470}\\}
\date{}
\begin{document}

\title{kappa-deformed quantum field theory and Casimir effect
}

\maketitle

\begin{abstract}
We consider the quantization of a scalar $\kappa$-deformed field up to the point
of obtaining
an expression for its vacuum energy. The expression is given by the half sum of
the field frequencies,
as in the non-deformed case, but with the frequencies obeying the
$\kappa$-deformed dispersion relation.
We consider a set of $\kappa$-deformed Maxwell equations and show that for
the purpose of calculating
the Casimir energy the Maxwell field, as in the non-deformed case, behaves
as a pair of scalar fields.
Those results provide a foundation for computing the Casimir energy starting
from the the half sum of field frequencies. A method of calculation
starting from this
expression is briefly described.
\vspace{1pc}
\end{abstract}

\maketitle

\section{Introduction}

The $\kappa$-deformed Poincar\'e algebra
\cite{Lukierski91,Lukierski92} is a quantum group, {\it i.e.}, a
non-commutative and non-cocommutative Hopf algebra,
\cite{Chaichian96} obtained by a Wigner-Inonu contraction of a
deformation of the anti-De Sitter algebra
\cite{Lukierski91,Lukierski92}. The deformation parameter $\kappa$
has dimension of mass and can be related to the Planck's length if
the $\kappa$-deformation is proposed to describe spacetime
symmetries at the Planck scale. In the limit in which
$\kappa\!\rightarrow\!\infty$ the deformation disappears and the
commutation relations of the algebraic part of the Hopf algebra
reduce to the commutation relations of the usual Poincar\'e
algebra. The coalgebraic part becomes cocommutative in this limit,
but will not be further mentioned because is of no concern to us
here. The $\kappa$-deformed Poincar\'e algebra has the same 10
generators of the Poincar\'e algebra: the ones for rotations,
boosts, space translations ${\bf P}$ and time translations $P^0$.
They obey deformed commutation relations\cite{Lukierski91} for
which the following first Casimir invariant can be obtained:
\begin{equation}\label{kappaCasimirinvariant}
\left(2\kappa\,\hbox{senh}\frac{P^0}{2\kappa}\right)^2-{\bf P}^2=m^2 \; ,
\end{equation}
where the scalar $m$ is endowed with the usual interpretation of the mass of the particle in each irreducible
representation. For us here, the $\kappa$-deformed mass shell condition and dispersion relation obtained from the
invariant (\ref{kappaCasimirinvariant}) is the essential ingredient of the $\kappa$-deformation of the Poincar\'e
algebra. In the limit $\kappa\!\rightarrow\!\infty$ in which the deformation desappears the Casimir
invariant (\ref{kappaCasimirinvariant}) reduces to the usual one in the Poincar\'e algebra, to wit:
$P_0^2-{\bf P}^2=m^2$.

A field theory is $\kappa$-deformed when its spacetime symmetry is
governed by the $\kappa$-deformed Poincar\'e algebra. Let us
notice that field theories with $\kappa$-deformation may be
treated in the context of non-commutative spacetime
\cite{Kosinski00,Kosinski01}, but here we follow the original
approach of a deformed Poincar\'e algebra in usual spacetime
\cite{Lukierski91,Lukierski92}. In a $\kappa$-deformed field
theory the equations of motion for the free fields should give
rise to the mass shell and dispersion relation given  by
(\ref{kappaCasimirinvariant}). Examples of such equations were
proposed for a $\kappa$-deformed scalar field \cite{Lukierski92},
\begin{equation}\label{kappaKG}
(\nabla^2-\partial_q^2-m^2)\,\phi({\bf x},x^0)=0 \;
\end{equation}
and the $\kappa$-deformed Dirac field \cite{Lukierski92}:
\begin{equation}\label{kappaDirac}
(i\gamma^i\partial_i+i\gamma^0\partial_q-m)\psi({\bf x},x^0)=0 \; ,
\end{equation}
where the gamma matrices are the usual ones and use was made of the definitions:
\begin{equation}\label{partialq}
q=\frac{1}{2\kappa}\;\;\hbox{and}\;\;\partial_q=\frac{1}{q}\,{\rm sen}{(q\,\partial_0)} \; .
\end{equation}
Let us notice that $\partial_q$ is a differential operator of infinite order, except in the limit
$q\!\rightarrow \! 0$ in which de deformation disappears and $\partial_q\!\rightarrow\! \partial_0$.
The infinite order of this operator is a main source of complexity in the calculations in deformed theories.

Another example of $\kappa$-deformed equations of motion is given by the following $\kappa$-deformed Maxwell
equations \cite{C-PFM02}:
\begin{eqnarray}
\nabla\cdot{\bf E}=0 \; ,\label{qMaxwell1}\\
\nabla\times{\bf E}=-\partial_q{\bf B} \; ,\label{qMaxwell2}\\
\nabla\cdot{\bf B}=0 \; ,\label{qMaxwell3}\\
\nabla\times{\bf B}=\partial_q{\bf E}\; .\label{qMaxwell4}
\end{eqnarray}
Notice that relativistic wave equations for free fields were
obtained by the Wigner construction in the context of
noncommutative $\kappa$-deformed spacetime \cite{Kosinski01}.
Returning to the $\kappa$-deformed Maxwell equations
(\ref{qMaxwell1}-\ref{qMaxwell4}) we obtain for their plane wave
solutions the dispersion relation
\begin{equation}\label{kappaemdispersion}
\sinh(q\,\omega)=q\,|{\bf k}| \; ,
\end{equation}
in accordance with the $\kappa$-deformed Casimir invariant (\ref{kappaCasimirinvariant}).

Let us now consider the Casimir effect. The original Casimir effect \cite{Casimir48} consists
in the attraction between two parallel perfectly
conducting plates at small separation. This attraction is a consequence of the shift in vacuum
energy of the electromagnetic field which occurs due to the presence of the plates. The shift in
energy properly regularized and renormalized is known as Casimir energy and is given by:
\begin{equation}\label{E(a)}
{\cal E}(a)=-\frac{\pi^2\ell^2}{720{a}^3}\; .
\end{equation}
where the plates are considered as two squares of side $\ell$ separated by a distance $a$
($a\ll\ell$). It is a QED phenomenum, although at the one-loop level, in which it is defined, the
quantized electron field can be ignored. In this case its only effect is described by the bulk action
of the perfectly conducting plates on the electromagnetic field through the boundary conditions.
In a broader sense the Casimir effect \cite{Casimir48} is a property of any relativistic quantum field
which arises from the shift in their vacuum energy due to boundary conditions or other properties
of the base manifold of the field \cite{Bordag01}). It is a fundamental effect and at the same time
requires for its calculation a relatively small part of the theory
of the field in consideration, namely: the part which leads to an expression for the vacuum energy or other
equivalent quantity. These two properties make the Casimir effect specially well suited for
the investigation of theories still under construction. The Casimir effect for $\kappa$-deformed
electrodynamics was calculated perturbatively in the parameter $1/\kappa$ \cite{Bowes96} and also
non-perturbatively \cite{C-PFM02}, from the sum of modes expression of the vacuum energy. It was also
obtained for a massive scalar field from the Schwinger proper time representatio of
the effective action \cite{C-PF97}. Here we address the question of taking for the vacuum energy in the
$\kappa$-deformed case the usual expression of half sum of the field frequencies. Notice that $\kappa$-deformation
may change drastically some quantities associated with the fields, as it happens with the one-loop effective
action \cite{C-PF97}.
\section{Vacuum energy in $\kappa$-deformed theories}

The Casimir effect is obtained from the vacuum energy, which results from quantum fluctuations and, as such,
is a fundamental quantity of any quantum field.  In the case of $\kappa$-deformed fields there is no
complete quantized theory yet. However, it is possible to bypass difficulties due to the deformation and
develope quantization to the point of obtaining the expression for the vacuum energy. We will consider now
the case of a scalar field for simplicity. More complicated kinematics will not affect the main features of
the formalism below.

We start from a Lagrangian which is equal up to surface terms to the one
proposed by Lukierski, Nowicki and Ruegg \cite{Lukierski92}:
\begin{equation}\label{L}
{\cal L}=\frac{1}{2}\partial_{\bar{\mu}}\phi\partial^{\bar{\mu}}\phi
-\frac{1}{2}m^2\phi^2 \; ,
\end{equation}
where we are using the convention that a bar over an index means that its range is
$\{q,1,2,3\}$. The action for this infinite order Lagrangian has the general form
\begin{equation}
{\cal W}(\phi)=\int_{\Omega}d^4x\;{\cal L}(\phi{(x)},\bar{\partial}\phi{(x)}) \; .
\end{equation}
where $\bar{\partial}\phi=(\partial_q\phi,\partial_1\phi,\partial_2\phi,\partial_3\phi)$
and $\Omega=\R^3\times[t_1,t_2]$. Taking a virtual variation of this action and using the lemma:
\begin{eqnarray}
\Upsilon(\partial_q\Xi)+ (\partial_q\!\!\!\!\!\!\!\!\!\!\! &&\Upsilon)\Xi=
\partial_0\Big[-\frac{1}{q}\sum_{n=0}^{\infty}
(-1)^n\frac{q^{2n+1}}{(2n+1)!}\nonumber\\
&&\sum_{p=1}^{2n+1}(-1)^p(\partial_0^{2n+1-p}\Upsilon\partial_0^{p-1}\Xi)\Big] \; ,
\end{eqnarray}
where $\Upsilon$ and $\Xi$ are functions of time, we obtain:
\begin{eqnarray}
\delta {\cal W}(\phi)\!\!\!\!\!\!\!\!\! &&=
\int_{\Omega}d^4x\; \Bigg[\frac{\partial{\cal L}}{\partial\phi}
-\partial_{\bar{\mu}}\left(\frac{\partial{\cal L}}{\partial(\partial_{\bar{\mu}}\phi)}\right)
\delta\phi\Bigg]\nonumber\\
&&+\int_{\Omega}d^4x\;
\partial_\mu(\Pi^\mu\delta\phi)\; ,\label{deltaW}
\end{eqnarray}
where
\begin{eqnarray}\label{Pi0}
\Pi^0=-
\frac{1}{q}\!\!\!\!\!\!\!\!\! &&\sum_{n=0}^{\infty}
(-1)^n\frac{q^{2n+1}}{(2n+1)!}\nonumber\\
&&\sum_{p=1}^{2n+1}(-1)^p\partial_0^{2n+1-p}\frac{\partial{\cal L}}{\partial(\partial_q\phi)}\partial_0^{p-1}
\end{eqnarray}
and
\begin{equation}\label{Pii}
\Pi^i={\partial{\cal L}}/{\partial(\partial_i\phi)} \; .
\end{equation}
The action principle requires that the variation (\ref{deltaW}) depends only on surface terms and so we obtain
in de $\kappa$-deformed case the following Euler-Lagrange equation:
\begin{equation}\label{Euler-Lagrange}
\frac{\partial{\cal L}}{\partial\phi}
-\partial_{\bar{\mu}}\frac{\partial{\cal L}}{\partial(\partial_{\bar{\mu}}\phi)}=0 \; .
\end{equation}
Substituing the Lagrangian (\ref{L}) in this equation we obtain the $\kappa$-deformed Klein-Gordon
equation (\ref{kappaKG}). This shows that the Lagrangian (\ref{L}) in fact describes the
Klein-Gordon field. Since it does not depend on time explicitly, we get from Noether's theorem the
following conserved energy-momentum tensor:
\begin{equation}\label{Tetamunu}
\Theta^{\mu\nu}=\Pi^\mu\partial^\nu\phi-g^{\mu\nu}{\cal L} \; ,
\end{equation}
where $\Pi^\mu$ is given by (\ref{Pi0}) and (\ref{Pii}). From this tensor we get for
the energy $P^0$ of the $\kappa$-deformed field the expression:
\begin{equation}\label{P0}
P^0={\int}d^3{\bf x}\;(\Pi^0\partial^0\phi
-{\cal L})
\end{equation}
where we should notice that the similarity with the energy of the non-deformed case is only notational.
In fact $\Pi^0$ is the differential operator (\ref{Pi0}) in the deformed case and it reduces to a field
quantity only in the limit $q\!\rightarrow\! 0$.

For the quantization of the system the first step is to promote $\phi(x)$ to an operator:
\begin{equation}\label{phi}
\phi(x)=\int d^3{\bf p}\,\,\eta({\bf p})\left[a({\bf p})\,e^{-ip{\cdot}x}+a^{\dagger}({\bf p})\,e^{ip{\cdot}x}\right],
\end{equation}
where $\eta({\bf p})$ is a normalization factor, the plane waves have dispersion relation
\begin{equation}\label{kappadispersion}
p^0=\omega({\bf p})=\frac{1}{q}\sinh\left(q\sqrt{{\bf p}^2+m^2} \right)
\end{equation}
and their amplitudes $a({\bf p})$ and $a^{\dagger}({\bf p})$ are operators obeying the canonical commutation
relations:
\begin{equation}\label{CCR1}
[a({\bf p}),a({{\bf p}^{\prime}})]=[a^{\dagger}({\bf p}),a^{\dagger}({{\bf p}^{\prime}})]=0
\end{equation}
\begin{equation}\label{CCR2}
[a({\bf p}),a^{\dagger}({{\bf p}^{\prime}})]=\delta{({\bf p}-{\bf p}^{\prime})}
\end{equation}

The second step in the quantization procedure is to postulate the Heisenberg equation:
\begin{equation}\label{Heisenberg}
\partial^0\phi=i[P^0,\phi] \; ,
\end{equation}
where $P^0$ is the operator obtained by quantization of the energy (\ref{P0}).
By substituting in this operator the expression (\ref{phi}) for the field and performing
a lengthy calculation we obtain:
\begin{eqnarray}\label{P0eta}
P^0=\int\!\!\!\!\!\!\!\!\! &&  {d}^3{\bf p}
\left[\eta({\bf p})^{2}2(2\pi)^3\frac{{\rm  senh}(2q\omega({\bf p}))}{2q}\right]
\nonumber\\
&&\frac{1}{2}\omega({\bf p})\left(a^{\dagger}({\bf  p})a({\bf p})+a({\bf p})a^{\dagger}({\bf p})\right).
\end{eqnarray}
The compatibility of the Heisenberg equation (\ref{Heisenberg}) with the normalization chosen for the
commutation relations (\ref{CCR1}) and (\ref{CCR2}) requires for the normalization factor $\eta({\bf p})$
that $\eta({\bf p})^2\;2(2\pi)^3 {{\rm  senh}(2q\omega({\bf p}))}/{2q}\!=\!\!1$. This condition can obvioulsy be
fulfilled and used in (\ref{P0eta}) to obtain the following simple expression for the energy
operator of the $\kappa$-deformed field:
\begin{equation}\label{P0operator}
P^0\!=\!\int{d}^3{\bf p}\,\frac{1}{2}\omega({\bf p})
\left(a^{\dagger}({\bf p})a({\bf p})+a({\bf p})a^{\dagger}({\bf
p})\right) ,
\end{equation}
where the frequency $\omega({\bf p})$ is given by the $\kappa$-deformed dispersion relation
(\ref{kappadispersion}). By using in (\ref{P0operator}) the commutation relation (\ref{CCR2})
and the definition $a({\bf p})|0\rangle=0$ for the vacuum state $|0\rangle$, we obtain for the
vacuum energy $E_0$ of the $\kappa$-deformed field in box normalization \cite{pc02n1.tex}: 
\begin{eqnarray}\label{E0}
E_0\!\!\!\!\!\!\!\!\! &&=\sum_{\bf p}\frac{1}{2}\omega({\bf p})\nonumber\\
&&=\sum_{\bf p}\frac{1}{2q}\sinh^{-1}(q\sqrt{{\bf p}^2+m^2})\; .
\end{eqnarray}
Therefore, by using canonical quantization procedures for the $\kappa$-deformed field
we obtain that its vacuum energy is given, as in the non-deformede case, by the half sum
of the field frequencies. In this way the vacuum energy of the deformed field differs from
the vacuum energy of the non-deformed only in that the dispersion relation for the frequencies of the
former is a deformation of the dispersion relation for the frequencies of the latter.

Now that we arrived at the expression for the vacuum energy ({\ref{E0}) of the $\kappa$-deformed field
two remarks are in order. The first is that the consistency of the canonical quantization requires that
the Heisenberg equation (\ref{Heisenberg}) and the commutation relations (\ref{CCR1})-(\ref{CCR2})
must lead to the same equation of motion obtained from the $\kappa$-deformed Euler-Lagrange equations
(\ref{Euler-Lagrange}), namely, to the $\kappa$-deformed Klein-Gordon equation (\ref{kappaKG}). This can be
verified by using for the differential operator $\partial_q$ in (\ref{partialq}) the factorization
${\partial}_{q}={\partial}_{q}^{\flat}\;\partial_0$, where
\begin{equation}
{\partial}_{q}^{\flat}=
\frac{1}{q}\sum_{n=0}^{\infty}\frac{(-1)^{n}q^{2n+1}}{(2n+1)!}\partial_{0}^{2n}\; ,
\label{DerondBarraQ}
\end{equation}
and applying the differential operator ${\partial}_{q}^{\flat}$ to both sides of the Heisenberg
equation (\ref{Heisenberg}). The second remark is that the quantization was performed without
using a conjugate field for the field $\phi$. This is why the canonical commutation relations
were imposed on the oscillators $a({\bf p})$ and $a^{\dagger}({\bf p})$, as given in (\ref{CCR1})
and $(\ref{CCR2})$. The choice of a conjugate field is far from trivial in the case of infinite order
Lagrangians, as it is the case of the $\kappa$-deformed Lagrangian (\ref{L}). In the deformed case there
are several quantities playing the role of conjugate fields. We have, for example, the variable:
\begin{equation}\label{piq}
\pi_{q}(x)=\frac{\partial\cal L}{\partial(\partial_q\phi)}=\partial_{q}\phi(x) \; ,
\end{equation}
which takes the place of the time derivatives of the field after a transformation of Legendre type
is performed on the Lagrangian. In order to obtain the equal time canonical commutation relations
involving this field we derive from (\ref{phi}), (\ref{CCR1})and $(\ref{CCR2})$ the following general
commutator for the field:
\begin{equation}
[\phi(x),\phi(x^\prime)]=i\,\Delta_q(x-x^\prime) \; ,
\end{equation}
where $\Delta_q$ is the Pauli-Jordan function of the $\kappa$-deformed field:
\begin{eqnarray}
\Delta_q(x-x^\prime)=\!\!\!\!\!\!\!\!\! &&\frac{-i}{(2\pi^3)}\int
\frac{d^3{\bf p}}{\sinh{2q\omega({\bf p}})/q}\nonumber\\
&&\left[
e^{-ip\cdot(x-x^\prime)}-e^{ip\cdot(x-x^\prime)}
\right] \; ,
\end{eqnarray}
which has the form proposed by Lukierski, Nowicki and Ruegg \cite{Lukierski92}.
From this general commutator we obtain that the Legendre conjugate field $\pi_q$
given in (\ref{piq}) obeys the commutation relation
\begin{eqnarray}
[\pi_{q}({\bf x},t),\phi({\bf x}^\prime,t),]\!\!\!\!\!\!\!\!\! &&=
i\,\partial_{q}\Delta_q(x-x^\prime)|_{x^0=x^{\prime\,0}}\nonumber\\
&&\not=-i\delta({\bf x}-{\bf x}^\prime) \; .
\end{eqnarray}
On the other hand, by using the field
\begin{equation}
\pi_{2q}(x)=\partial_{2q}\phi(x) \; ,
\end{equation}
we obtain:
\begin{equation}
[\phi({\bf x},t),\phi({\bf x}^\prime,t)]=0=
[\pi_{2q}({\bf x},t),\pi_{2q}({\bf x}^\prime,t)]
\end{equation}
and
\begin{equation}
[\phi({\bf x},t),\pi_{2q}({\bf x}^\prime,t)]= i\,\delta({\bf x}-{\bf x}^\prime) \; .
\end{equation}
Finally, we have in (\ref{P0}) the differential operator $\Pi^0$ playing the role of the
$\kappa$-deformed version of the conjugate field $\pi(x)=\partial_0\phi(x)$ in the
non-deformed case. In this way we have in the $\kappa$-deformed case the three objects
$\pi_q$, $\pi_{2q}$ and $\Pi^0$ exhibiting different properties of the conjugate field
$\pi(x)=\partial_0\phi(x)$ in the non-deformed case. All of them have the same expected
limit when the deformation disappears:
\begin{equation}
\lim_{q\rightarrow 0}\pi_{2q}=\lim_{q\rightarrow 0}\pi_{q}=\lim_{q\rightarrow 0}\Pi^0=\;\partial_0\phi\; .
\end{equation}
The important point is that it was not necessary to make a choice between those three quantities in
order to arrive at the expression of the vacuum energy in (\ref{E0}).
\section{The $\kappa$-deformed Casimir effect}

Let us now consider the Casimir effect in the case of $\kappa$-deformed electrodynamics.
As in the original Casimir effect we consider two parallel perfectly conducting squares of
side $\ell$ and separated by a distance $a$ ($a\ll\ell$). Those two plates are now in the
vacuum of the $\kappa$-deformed quantum electromagnetic field. To describe the effect of
the plates on this field we must consider the
inhomogeneous Maxwell equations with source terms added to equations (\ref{qMaxwell1}) and (\ref{qMaxwell4}).
It is easy to verify that those inhomogeneous equations give rise to the same boundary conditions
and transverse polarization properties that are obtained in the non-deformed case. As a consequence
the wave numbers describing the ressonant modes of the electromagnetic fields between the
perfectly conducting plates are not modified by the deformation. This result is used in the
expression for the electromagnetic version of (\ref{E0}), which is given by
\begin{eqnarray}\label{emE0}
E_0\!\!\!\!\!\!\!\!\! &&=\sum_{{\bf k},\sigma}\frac{1}{2q}\sinh^{-1}(q\,|{\bf k}|)\; ,
\end{eqnarray}
where $\sigma$ is the index for the transverse polarizations and the dispersion relation
(\ref{kappaemdispersion}) has been used. The resulting expression for the vacuum energy in the
presence of the plates is:
\begin{eqnarray}\label{E0em}
E_0=\frac{\ell^2}{2\pi{q}}
\sum_{n=1}^{\infty}\int_{0}^{\Lambda_\parallel}\!\!\!\!\!\!\!\!\! &&
{dk_\parallel}{k_\parallel}
e^{-\epsilon\sqrt{k_{\parallel}^2+k_n^2}}\nonumber\\
&&\sinh^{-1}\left(q\sqrt{k_{\parallel}^2+k_n^2}\right)
\end{eqnarray}
where $k_n=\pi n/a$ ($n=0,1,...$), $\epsilon$ and $\Lambda_{||}$ are positive regularization parameters
and the mode $n=0$ was ignored because it gives rise to the spurious term representing the
self-energy of the plates. After the required subtractions the regularization parameters are
eliminated by the limits $\epsilon\!\rightarrow\! 0$ and $\Lambda_{||}\!\rightarrow\!\infty$.
The expression (\ref{E0em}) provides the starting point of the calculation of the
Casimir energy in $\kappa$-deformed electrodynamics from the expression of vacuum energy
as a sum of modes \cite{Bowes96,C-PFM02}. Let us briefly describe how to obtain an expression of this
energy which is non-perturbative in the deformation parameter \cite{C-PFM02}. First of
all we notice that although the
cutoff $\Lambda_{||}$ may seem unnecessary in view of the regularization performed by the
parameter $\epsilon$, it must be used in order to apply the argument principle in the
calculation of the series in (\ref{E0em}). After the application of this principle it is
easy to identify in the resulting expression the spurious term describing the vacuum
energy in the absence of the plates. By subtracting this term we arrive at the Casimir energy
${\cal E}_q(a)$ that can be written as:
\begin{eqnarray}\label{Eq(a)}
{\cal E}_q(a)=
-\frac{\ell^2}{4\pi^2{a}^3}\sum_{n=1}^{\infty}\frac{1}{n^2}
\int_{0}^{a/q}\!\!\!\!\!\!\!\!\!\!\!\!\!\! &&
dy\left(y+\frac{1}{2n}\right) \nonumber\\
&&\frac{e^{-2ny}}{\sqrt{1-(qy/a)^2}} \; .
\end{eqnarray}
In the limit $q\!\rightarrow\! 0$ this energy reduces to the usual Casimir energy (\ref{E(a)}),
as expected.

\end{document}